# Widespread occurrence of lava lakes on Io observed from Juno

Alessandro Mura[1], Federico Tosi[1], Francesca Zambon[1], Rosaly M. C. Lopes[2], Pete J. Mouginis-Mark[3], Jani Radebaugh[4], Alberto Adriani[1], Scott Bolton[5], Julie Rathbun[6], Andrea Cicchetti[1], Davide Grassi[1], Raffaella Noschese[1], Giuseppe Piccioni[1], Christina Plainaki[7], Roberto Sordini[1], Giuseppe Sindoni[7]

1 Istituto Nazionale di Astrofisica – Istituto di Astrofisica e Planetologia Spaziali, Rome, Italy

2 Jet Propulsion Laboratory, California Institute of Technology, Pasadena, CA, USA

3 Hawaii Institute Geophysics and Planetology, Honolulu, HI, USA

4 Brigham Young University, Provo, UT, USA

5 Southwest Research Institute, San Antonio, TX, USA

6 Cornell University, Ithaca, NY, USA

7 Agenzia Spaziale Italiana, Rome, Italy

**Abstract.** We report recent observations of lava lakes within patera on Io made by the JIRAM imager/spectrometer on board the Juno spacecraft, taken during close observation occurred in the extended mission. At least 40 lava lakes have been identified from JIRAM observations. The majority (>50%) of paterae have elevated thermal signatures when imaged at sufficiently high spatial resolution (a few km/pixel), implying that lava lakes are ubiquitous on Io. The annular width of the spattering region around the margins, a characteristic of lava lakes, is of the order of few meters to tens of meters, the diameter of the observed lava lakes ranges from 10 to 100 km. The thickness of the crust in the center of some lava lakes is of the order of 5-10 m; we estimate that this crust is a few years old. Also, the bulk of the thermal emission comes from the much larger crust and not from the smaller exposed lava, so the total power output cannot be calculated from the 5-μm radiance alone. Eight of the proposed lava lakes have never been reported previously as active hotspots.

# Introduction

Io, the innermost of Jupiter's Galilean moons, stands out as one of the most intriguing bodies in the Solar System due to its extraordinary volcanic activity. First observed in detail by the Voyager missions in the late 1970s, Io was quickly recognized as the most volcanically active object in the Solar System, with its surface constantly reshaped by intense and frequent eruptions (Smith et al., 1979; Morabito et al., 1979). Unlike any other moon or planet, Io's dynamic geology is driven by the intense tidal forces exerted by Jupiter, which generate immense internal heat through frictional processes (Peale et al., 1979). This heat makes Io the only place, besides Earth, where widespread, active silicate volcanism has been observed (e.g., McEwen et al., 2000; Lopes et al., 2018), making it a natural laboratory for studying planetary volcanism and the extreme conditions that shape such bodies.

The predominant volcanic structures on Io are "paterae", which resemble terrestrial calderas but can be significantly larger, with some, like Loki, reaching to just above 200 km in diameter (Radebaugh et al., 2001). Thermal emissions are dominantly confined to the more than 400 paterae distributed globally. Additionally, Io features extensive lava flows across surface plains such as at Amirani volcano, which has an active lava flow over 300 km long (Williams et al., 2011a, 2011b). Some flows overflow out of paterae and elsewhere as standalone, fissure-fed flows, and which are also the source of thermal emissions.

The Galileo spacecraft, which orbited Jupiter from 1995 to 2003, offered critical insights into Io's volcanic processes, revealing that much of the moon's surface is covered with vast lava flows and large volcanic depressions known as paterae (McEwen et al., 2000; Radebaugh et al., 2001). Infrared observations from Galileo identified many hotspots on Io, suggesting the presence of intense and widespread volcanic eruptions (e.g. Lopes et al., 1999). The confined nature of these thermal emissions over time, placing them within volcanic paterae, revealed ongoing eruptions within these depressions. Furthermore, the thermal signatures revealed many patera eruptions were high temperature and high thermal output consistently over time, which implied they were lava lakes (Lopes et al. 2004). Lava lakes are unique features on Earth that require a direct connection to the interior and a magma source, and thus they can reveal aspects of lava properties and temperatures in interior source regions (Swanson et al. 1979; Harris et al. 1999). Lava lakes often have cool crusts of varying thickness because of radiative or convective cooling of the lava lake surface balanced with input of lava and gases from below (Flynn et al., 1993; Lev et al. 2019). Even when temperatures are high and convection rapid, at least some crust forms – the Marum lava lake of Vanuatu had convection speeds of ~5 m/s, and yet had cool crustal portions on the surface and at the margins (Radebaugh et al. 2016). This style of lake surface is termed "Chaotic", while a more continuous lake surface occurs on lava lakes termed "Organized" (Lev et al. 2019), such as the Erta Ale lava lake of Ethiopia and the Pu'u 'O'o lake of Kilauea, Hawaii (Figure 1S in supplementary material). At organized lava lakes cracks occur on the lake surface and especially at the lake margins, which act to expose the hot lava. Convection, fountaining, and gas emissions break apart the lava lake crust, and the crust in turn acts to help regulate the convection and outgassing (Lev et al. 2019).

To date, over 250 active volcanic features have been identified on Io (Lopes and Spencer, 2007; Veeder et al., 2015; Zambon et al., 2023; de Kleer and Rathbun, 2023), identified as having erupted at some point while we have been able to observe it, though many of these eruptions are ongoing. Although the exact composition of Ionian lavas has not been directly measured, inferred lava temperatures, colors and eruption edifice morphologies suggest they are likely basaltic or possibly ultramafic (Stansberry et al., 1997; McEwen et al., 1998; Allen et al., 2013; Keszthelyi and Suer, 2023). The precise nature of the subsurface magma plumbing system that supplies these lava lakes also remains unclear, with various models proposing different magma supply rates and corresponding heat flow (McEwen et al., 2004, Hamilton et al. 2013).

The Juno mission, which has been orbiting Jupiter since 2016, has further expanded our understanding of Io, particularly through its recent close flybys. Juno's infrared imaging has revealed new details about the temperature distribution and thermal emissions of Io's volcanic regions, providing fresh data on the mechanisms driving Io's volcanic activity (Mura et al., 2020; Zambon et al., 2023; Mura et al., 2024a, 2024b). These observations are crucial for developing more accurate models of Io's internal structure and the processes that sustain its volcanic activity.

This paper aims to build on our previous work by analyzing the most recent data from Juno's observations of Io, with a focus on mapping the lava lakes that can be identified by the observations by JIRAM. By integrating these new findings with established knowledge, we seek to provide a comprehensive view of Io's ongoing volcanic activity and its underlying causes.

## Instrument and data set.

JIRAM (Adriani et al., 2017) is an infrared imager and spectrometer on board the NASA Juno spacecraft. The imager is a single detector subdivided into two filter subsections (L band: 3.3 to 3.6 µm; M band: 4.5 to 5 µm). The spectrometer covers the range from 2 to 5 µm, but this channel is not used in this work; more instrument details pertinent to Io's observations can be found in Mura et al., 2020. During each Juno rotation, JIRAM captures two 2D images, one in the L and one in the M bands. The boresight of JIRAM's Field of View (FoV) is within the plane perpendicular to the Juno spin axis, so the acquisition timing defines the direction of the observation. FoV adjustments in other directions require spacecraft reorientation, sometimes performed for observing Io. A de-spinning mirror was designed to compensate for the spacecraft's rotation (about 2 rpm), but due to an electronic board malfunction, the mirror has not been used since orbit 44. For high-radiance targets like hotspots, the short exposure time (~few milliseconds) minimizes image smearing, which is corrected via software. The pixel angular resolution (IFOV) is 237.7 µrad, providing a FoV of 5.87° by 1.74° for both the L and M bands; the NER (Noise Equivalent Radiance) depends on the integration time, and is below 0.5 mW sr$^{-1}$ m$^{-2}$ at 10 ms.

JIRAM took almost 2000 M-band and L-band images of Io during Juno's orbits 51, 53, 55, 57, 58, 60 and 62. Two close approaches/flybys occurred on orbits 57 and 58, when the spatial resolution reached 300 m. From orbit 60 onwards, the distance between the orbits of Juno and Io has been increasing, so this is the best resolution JIRAM ever achieved. The typical spatial resolution of our dataset is in the range from 1 to 10 km per pixel. We note that, due to the extremely large dynamics of the radiance from the hotspots of Io, the JIRAM data can be saturated in few cases, even when the exposure time is a few ms (as in this dataset). The analysis presented here is mostly based on the overall thermal distribution within the hot spot, so the few saturated images do not compromise our conclusions.

In this study we analyze 61 hotspots observed by JIRAM, starting with the hypothesis that many could be lava lakes, as discussed above and reported in our previous study (Mura et al., 2024b). In that paper, we focused on a few hotspots with evident ring-type emission around the edges. Here we broaden our data set by using observations with better spatial resolution (orbits 53, 55, 57, 58, 60 and 62), down to 300 m. In total we have 140 radiance maps (for many hotspots we have more than one observation, possibly in both bands); some of these maps are obtained as a "super resolution" map as in Mura et al. (2020) or Mura et al. (2024b) (i.e. stacking more images of the same region, co-registered, and using the median value to increase the quality of the data); in some cases, the hotspot is seen only in one observation and only one map is available.

## Distribution of lava lakes

We analyzed the images taken by JIRAM during its orbits 51, 53, 55, 57, 60 and 62. We focused on all features with a well-defined morphology, i.e., thermal emission with a shape different from a single and uniform spot concentrated on a few pixels (basically, everything than what is plausibly a sub-pixel hotspot, which in principle should be concentrated in a single pixel but may form a compact cluster of several pixels without specific shape due to JIRAM's smearing and point spread function). In Figure 1 we show four examples of the hotspots/lava lakes analysed in this study. The level of detail present in these images of thermal emissions on Io is unprecedented except in a few high-resolution Galileo images of Pele and another region (McEwen et al. 2000; Radebaugh et al. 2004). Note the continuous, bright "hot rings" that correlate with the outlines of paterae wall as seen in Galileo and Voyager images. This indicates the rings could be exposed lava at the margins of a cooled crust on a potential lava lake contained inside the paterae, corroborating previous conclusions about these features. Note the resolution differences in the different images of Figure 1; Catha P. map in panel A, has a resolution of 9 km/pixel, and thus the ring appears broad, while Chors P., in panel B, has a resolution of 3 km/pixel and shows a slightly thinner ring. In both cases, the ring is quite continuous around the margin. In panels C and D, the resolution of the map is ~500 m/pixel, and the rings appear much thinner. Also notable in panel C is the nearly continuous nature of the margin, as well as the consistency in thickness. The meandering nature of the line is also unique and is an improvement on the resolution of the shape of the patera margin. The margin of the radiance map in panel D is unique and challenging to interpret, also because there is no currently available visible counterpart; it could represent a nested patera, or perhaps a patera with an elevated, cool island, which is also common. Because of the large number of hotspots, all other ones are reported, with images and a few comments, in the supplementary material.

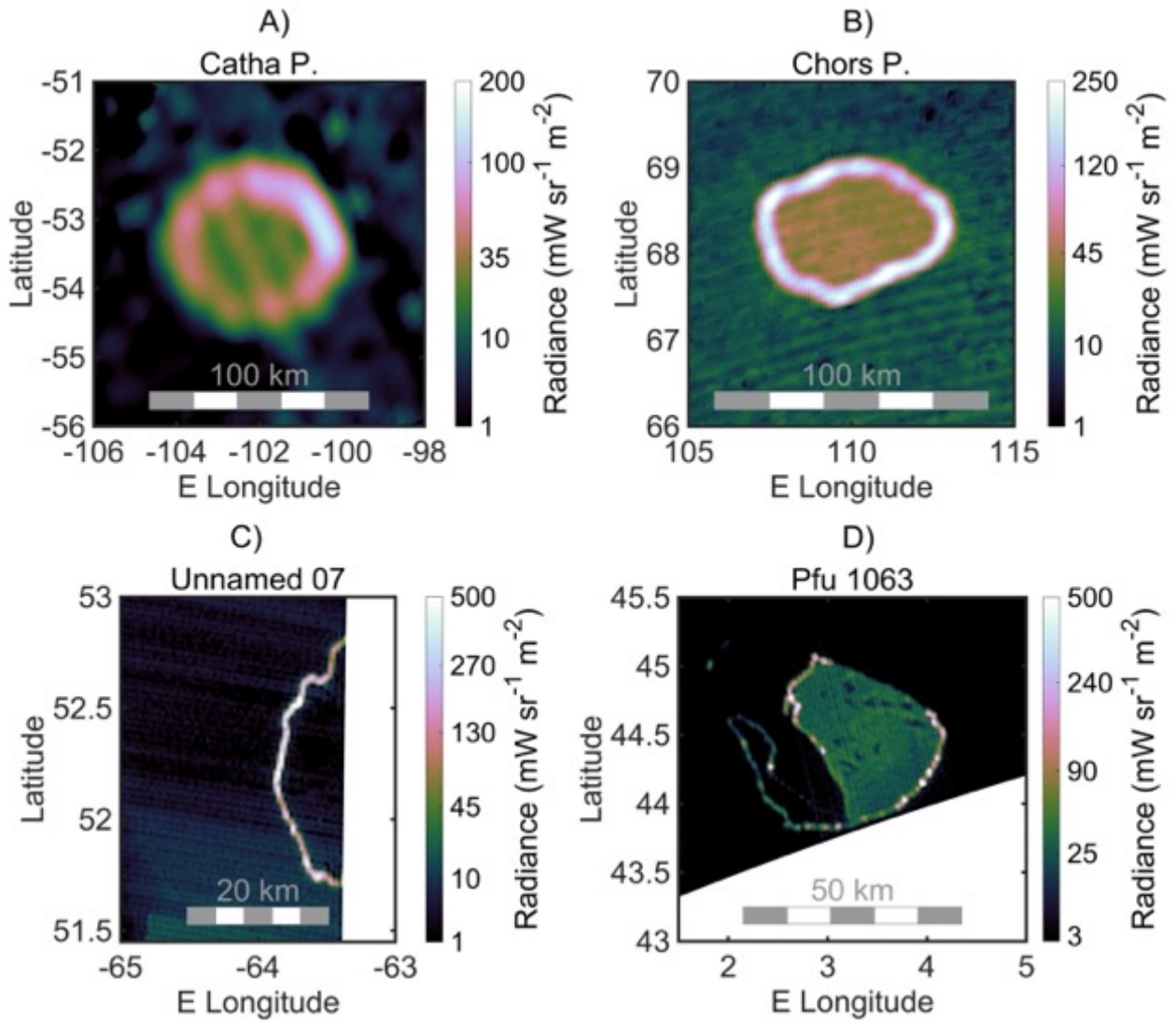

**Figure 1.** Four different examples of lava lakes, in order of increasing spatial resolution. Catha Patera (panel A), Chors Patera (panel B), Unnamed 07 (panel C, only partially visible) and an unnamed hotspot previously designated Pfu1063 by Cantrall et al., (2018) (panel D). Color scales are not linear, and the limits are adjusted to enhance the low-radiance regions.

At present, no currently approved space mission will observe Io with a spatial resolution better than these images, which means that more detailed measurements than these will not be available for a long time. We also note that, despite a large quantity of data that covers Loki Patera, we exclude it from this analysis as it is a unique case within Io's volcanic phenomena, and as such it deserves a future, separate study. With the assumption that the hot rings in the IR images are located at the perimeters of a lava lakes (Mura et al., 2024a), an algorithm is used to locate the perimeter and calculate the area. Most features can be fitted by this algorithm; those that can be attributed to a complex morphology are relatively few, such as Amirani (a flow), which cannot be easily associated with a lava lake or a lava lake system at these resolutions. It must also be noted that the fitted perimeter depends on the spatial resolution of the data. In some case such as Chors, a patera that appears circular, crenulated edges are visible when the resolution increases (from orbit 51 to orbit 55, resolution increases from 10 km/pixel to 3 km/pixel; see figure 5 in Mura et al., 2024b).

The distribution of the features that we interpret as lava lakes appears to be scattered evenly across the surface of Io; there may be a peak in distribution slightly offset from the sub- and anti-Jovian hemispheres, at 310 W and 135 W, similar to the distribution of paterae noted in previous work (Radebaugh et al. 2001; Hamilton et al. 2013); however, the spatial resolution is not uniform over the surface of Io, so that further analysis is needed to confirm this result. Finally, the distribution of the circularity factor $f$ (see methods) does not show that more circular lava lakes are concentrated in any preferred region. There is no clear dependence of $f$ with latitude or longitude, nor does it peak at any specific location. But since our coverage and spatial resolution are not uniform over Io's surface, we only preliminarily conclude that the size or morphology of the observed lava lakes is variable, on Io, in a stochastic way.

It is sometimes difficult to determine when the thermal distribution within a hotspot has a ring shape because, for example, paterae completely filled with lava can appear similar to unresolved hotspots. However, since it is quite clear that the majority of paterae show thermal rings, we can say with reasonable certainty that the fraction of hotspots that turn out to be lava lakes with ring-like features is 50% or more, at least for those that can be detected with a resolution of 1-10 km (about 40 out of 60 features observed). Very small hotspots cannot be resolved because of the limit in the spatial resolution. Since the dominant structure of the larger hotspots could be systematically different from the smaller ones, we cannot draw conclusions about hotspots for which we cannot resolve the thermal structure, even when the resolution achieves 300 m (JIRAM's highest resolution; most of Io's surface, however, is not covered at this spatial resolution). For these, a more precise comparison can be made in the future using JunoCam images.

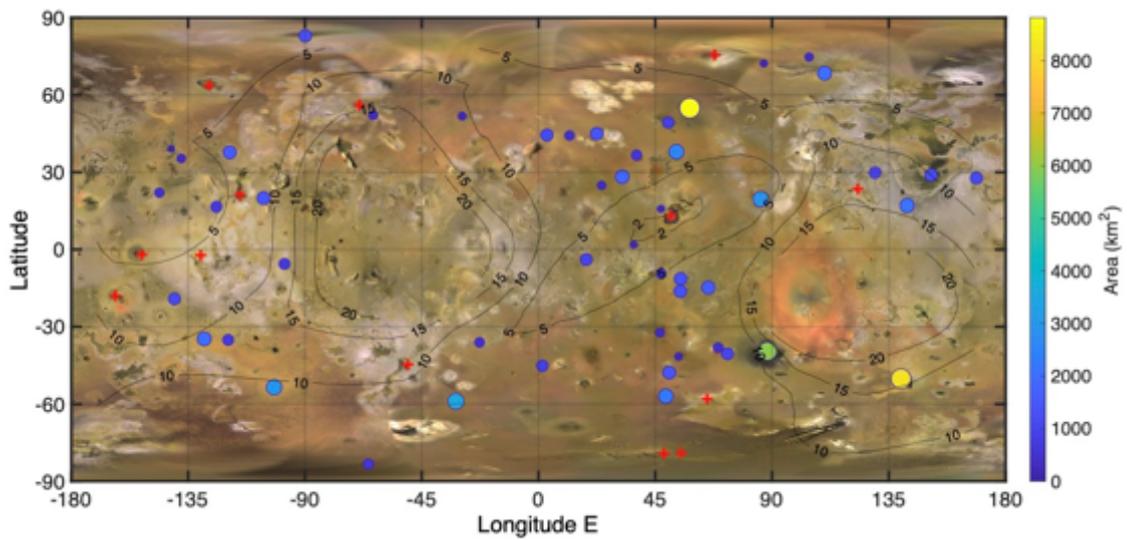

**Figure 2:** Position of lava lakes, indicated by circles. The areas of the circles are proportional to the areas of the lakes (which range from 100 km$^2$ to 9000 km$^2$), colors are coded according to the color bar to the right. Red crosses are flows or other features. Contour lines indicates the spatial resolution (best case), in km, obtained by JIRAM while imaging the surface of Io. Orbits 57 and 58 have very low coverage and are not visible in the contour plot.

We also note that the shape of these hot rings varies from circular to irregular (though shape is not quantified in this initial report, except for the circularity factor $f$). Among the 61 features we analyze, nine hotspots have an emission pattern that is very close to a perfect circular ring, and an inner crust that is featureless: Babbar P., Catha P., Dazhbog P., Maui P., Mulungu P., the unnamed hotspot previously designated as P63 in Veeder et al. (2015), Sigurd P., Surt P., and a previously unknown active patera (which we designate U08). Thirty-three hotspots have an emission that shows a non-circular ring, or a part of the ring that is clearly more intense regardless of the viewing geometry, or that the crust shows some particular morphology, or that there are flows outside the patera walls: Amaterasu P., Aten P., Chors P., Daedalus P., Dusura P., Fuchi P., Gibil P., Heno P., Kibero P., Kinich Ahau P., Malik P., Mazda P., Mihr P., Paive P., the hotspots known as P62, P151, PFd1069, PFd630, PV59 and PV92 in Veeder et al. (2015), the hotspot known as Pfu1063 in Cantrall et al. (2018), Tol Ava P., Ulgen P., the unnamed hotspot in Bulicame Regio P., Uta P., the hotspot designated as Unnamed 1 (U01) in Mura et al., 2024b, and 7 unnamed hotspots (which we designate U02, U04, U05, U06, U07, U17, U18). However, U04 can be safely placed inside Dingir Patera, and U17 is Hatchawa patera; they were not reported to be active hotspots before. U02, U05 and U06 are likely JR055, JR230 and JR229 in Zambon et al. (2023), respectively. The above-mentioned hotspots can be assumed to be lava lakes because of their thermal structure. Nineteen hotspots show very irregular emission patterns; some can still be interpreted to be lava lakes based on the comparison between infrared and visible images. In other cases, a lava lake appears to be only part of a larger volcanic feature, or the whole hotspot appears to be a flow, or else the morphology is too complex to be understood. These are: Amirani, Culann NW, Girru P., Isum P., Kurdalagon P., Manua P., Masubi F., Monan P., unnamed features known as PV129 in Veeder et al. (2015), Prometheus, Rarog P., Seth P., Tupan P., Tvashtar B-C, Vivasvant P., and 4 unnamed features (which we designate U13, U14, U15, U16; U16 is likely JR207 in Zambon et al., 2023). For some of them, if the presence of a lake is proposed, possibly in one part of the feature, the fitting algorithm is run, and they are plotted in Figure 2; remaining features are excluded from the analysis and are reported with only latitude and longitude in Table 1 (gray part of the table).

For each of these hotspots, the best available image is provided in the supplementary material, along with comments about notable features of the hotspot. Three other peculiar examples of lava lakes are shown Figure 3: Amaterasu Patera in two different phases of activity (panels A and B); Uta Patera, a boiling or chaotic lava lake (panels C and D); Tupan Patera, for which the morphology of the IR map is not easy to interpret (panels E and F).

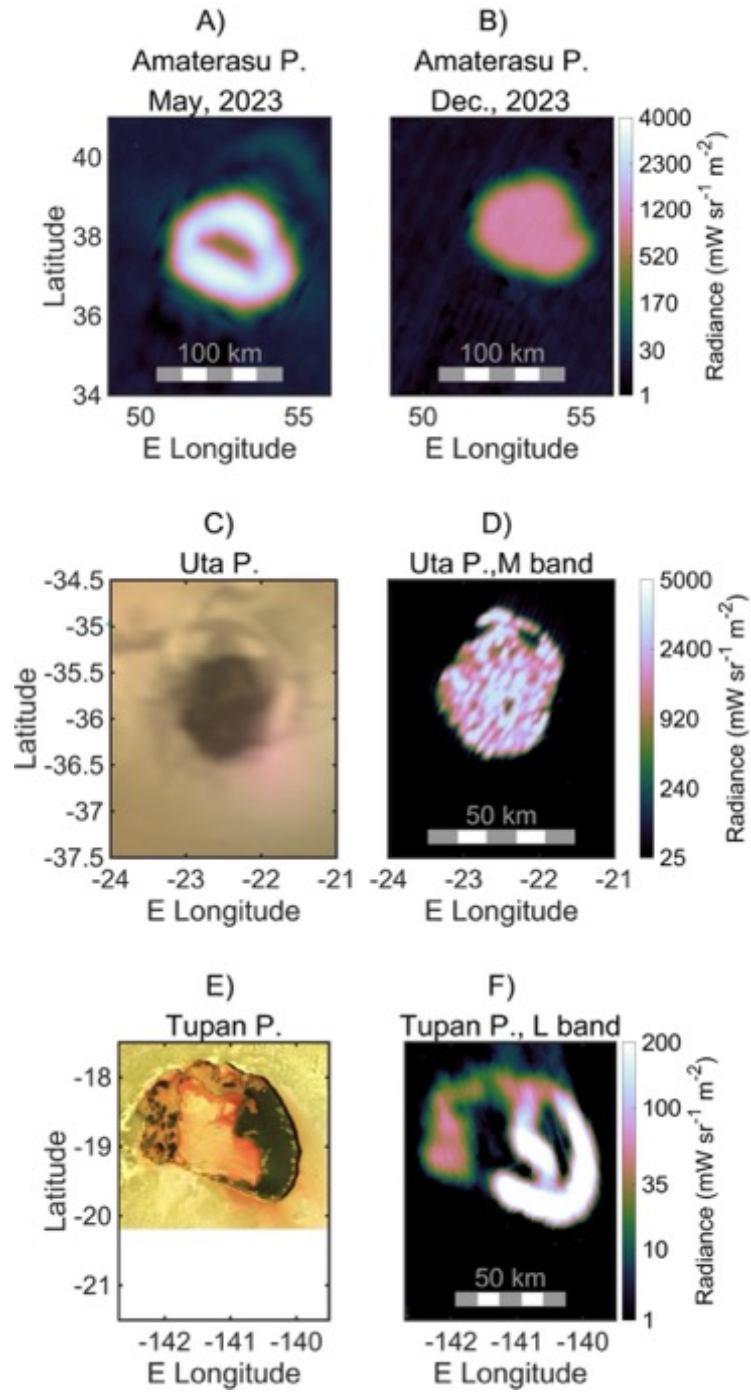

**Figure 3.** Panels A and B show Amaterasu P. at two different epochs, revealing the high variability of both the morphology and the intensity of the IR radiance; this was not an effect of the viewing geometry, which is at low emission angles in both observations. Panels C and D show the VIS and IR maps of Uta P., a boiling / Chaotic type of lake. Panels E and F show the VIS and IR maps of Tupan P.; the stripe north of the patera in the IR is an artifact. In the visible (from Galileo, credit NASA/JPL/University of Arizona), Tupan Patera is 80 km across, and the source of thermal emission is in the eastern portion, where dark flows are located. A bright, cool island exists in the patera center, and evidence of active volatile emission can be seen in the red coloration and flow-like textures on the eastern floor. Note the thin, dark line surrounding the island and patera margin; such a feature might have appeared as a bright ring when it was being deposited through overflow and drain-back, disjointed by the island in this case, perhaps similar to Figure 1, panel D, if seen by JIRAM. In the IR, thermal emission in the eastern part is the brightest.

| Name | Longitude E (°) | Latitude (°) | Area (km²) | Perimeter (km) | Factor | M-band Power (GW) |
|------|-----------------|--------------|------------|----------------|--------|-------------------|
| Amaterasu P. | 53 | 38 | 2565 | 183 | 0.649 | 9.6 |
| Aten P. | 50 | -48 | 1481 | 141 | 0.575 | 0.2 |
| Babbar P. | 88 | -39 | 6039 | 276 | 0.63 | 0.9 |
| Catha P. | -102 | -53 | 3020 | 193 | 0.849 | 0.8 |
| Chors P. | 110 | 68 | 2052 | 163 | 0.645 | 0.8 |
| Daedalus P. | 86 | 20 | 2934 | 197 | 0.673 | 1.3 |
| Dazhbog P. | 58 | 55 | 8819 | 332 | 0.798 | 1.2 |
| Dusura P. | -119 | 38 | 1521 | 138 | 0.818 | 0.6 |
| Fuchi P. | 32 | 28 | 1878 | 155 | 0.665 | 0.9 |
| Gibil P. | 65 | -15 | 1629 | 142 | 0.794 | 0.3 |
| Heno P. | 49 | -57 | 2338 | 171 | 0.747 | 1.7 |
| Isum P. | 151 | 29 | 1360 | 134 | 0.548 | 3.9 |
| Kibero P. | 55 | -11 | 1365 | 138 | 0.535 | 0.1 |
| Kinich Ahau P. | 50 | 49 | 933 | 111 | 0.623 | 1 |
| Kurdalagon P. | 140 | -50 | 8147 | 338 | 0.491 | 12.4 |
| Malik P. | -129 | -35 | 2020 | 165 | 0.567 | 2.4 |
| Manua P. | 38 | 37 | 640 | 90 | 0.724 | 0.6 |
| Maui P. | -124 | 17 | 804 | 102 | 0.749 | 0.2 |
| Mazda P. | 47 | -9 | 522 | 82 | 0.503 | 0.2 |
| Mihr P. | 55 | -16 | 1372 | 131 | 0.757 | 0.8 |
| Monan P. | -106 | 20 | 1515 | 149 | 0.407 | 0.7 |
| Mulungu P. | 142 | 17 | 2172 | 166 | 0.738 | 0.7 |
| P151 | -142 | 39 | 102 | 37 | 0.527 | 0.1 |
| P62 | -146 | 22 | 420 | 73 | 0.744 | 0.3 |
| P63 | -138 | 35 | 300 | 61 | 0.804 | 0.2 |
| PFd1069 | 12 | 44 | 453 | 77 | 0.617 | 0.2 |
| PFd630 | -90 | 83 | 1318 | 130 | 0.726 | 0.1 |
| PV129 | 24 | 25 | 297 | 69 | 0.29 | |
| PV59 | 69 | -38 | 526 | 95 | 0.252 | 0.9 |
| PV92 | 47 | 16 | 140 | 49 | 0.245 | 0.1 |
| Paive P. | 1 | -45 | 909 | 147 | 0.071 | 0.7 |
| Pfu1063 | 3 | 44 | 1207 | 124 | 0.544 | 0.1 |
| Rarog P | 54 | -42 | 198 | 51 | 0.616 | 0.4 |
| Sigurd P. | -98 | -6 | 753 | 100 | 0.614 | |
| Surt P. | 22 | 45 | 1387 | 131 | 0.861 | 0.4 |
| Tol Ava P. | 37 | 2 | 147 | 45 | 0.548 | 0.1 |
| Tupan P. | -140 | -19 | 1119 | 135 | 0.315 | 2.9 |
| Ulgen P. | 73 | -40 | 1088 | 116 | 0.677 | 3.2 |
| Unnamed 01 | 130 | 30 | 984 | 118 | 0.484 | 0.7 |
| Unnamed 02 / JR055 | 47 | -32 | 344 | 70 | 0.443 | 0.1 |
| **Dingir P. (Unnamed 04)** | 18 | -4 | 1071 | 118 | 0.608 | 0.02 |
| Unnamed 05 / JR230 | 87 | 72 | 162 | 46 | 0.645 | 0.1 |
| Unnamed 06 / JR229 | 104 | 75 | 280 | 59 | 0.78 | 0.1 |
| **Unnamed 07** | -64 | 52 | 389 | 35 | 0.107 | 0.03 |
| **Unnamed 08** | -29 | 52 | 231 | 54 | 0.646 | |
| **Unnamed 15** | -120 | -35 | 748 | 98 | 0.734 | |
| **Hatchawa P. (Unnamed 17)** | -32 | -59 | 3504 | 219 | 0.483 | 0.3 |
| **Unnamed 18** | -66 | -83 | 631 | 94 | 0.448 | 0.4 |
| Unnamed Bulicame R. | 169 | 28 | 1157 | 120 | 0.782 | 0.3 |
| Uta P. | -23 | -36 | 484 | 92 | 0.404 | 11 |
| Amirani | -115 | 21 | | | | |
| Culann NW | -163 | -18 | | | | |
| Girru P. | 123 | 23 | | | | |
| Masubi F. | -50 | -45 | | | | |
| Prometheus | -153 | -2 | | | | |
| Seth P. | -130 | -2 | | | | |
| Tvashtar B C | -127 | 64 | | | | |
| **Unnamed 13** | 55 | -79 | | | | |
| **Unnamed 14** | 65 | -58 | | | | |
| Unnamed 16 / JR207 | -69 | 56 | | | | |
| Vivasvant P. | 68 | 76 | | | | |

Table 1: location, area, perimeter, *f* factor (for the features where a ring has been identified), and power in the M band for the observed features in alphabetical order; for some features, M-band power has not been estimated. Lava flows or other features that are not proposed to be lava lakes are in gray in the second part of the table. Features that haven't been reported as active hotspots before are in bold.

# Estimation of crust thickness

There are some cases in which the crust at the center of the lava lake is seen with sufficient spatial resolution and shows uniform radiance. This means that JIRAM is measuring the average radiance of the crust in an area far from the edges and free from contamination by hot-ring pixels. The cases we consider are Surt Patera in orbit 55 (20 mW sr$^{-1}$ m$^{-2}$), Catha Patera in orbit 62 (between 20 and 30 mW sr$^{-1}$ m$^{-2}$), Chors Patera in orbit 55 (40 mW sr$^{-1}$ m$^{-2}$), and U07, for which the crust is practically dark (below JIRAM's Noise-Equivalent Radiance, NER). Three of these cases are in Figure 1, and Surt Patera can be found in the supplementary material. For these four cases, we assume that what we see is a black body at a temperature equal to the measured brightness temperature. The emissivity for crust at low temperature is very close to 1. The estimated temperature values in the first 3 cases are respectively ~215 K, ~220 K, and ~230 K. The total radiant emittance $M_e$ (across the entire spectrum) for black bodies at these temperatures are 120 W m$^{-2}$, 130 W m$^{-2}$, and 150 W m$^{-2}$.

The following calculation can be considered a simplified version of the model in de Kleer and de Pater (2017). We discard the temporal variability, which is not needed since we can use the instantaneous values for the crust temperature; latent heat released by crystallization is not considered (the crust is old, the crystallization rate is low, and the latent heat is small). De Kleer and de Pater (2017) assume that liquid basaltic magma has a temperature of 1475 K, which is near to a temperature found for the Pele volcano using Galileo and Cassini images (Radebaugh et al. 2004); for basaltic lava flows made mostly of silica, Ebert et al. (2002) gives a lower value of about 1000 K (see their figure 7). In all three cases that we propose, the temperature difference between the magma and the outer surface of the crust can be estimated to be around 1000 K as an order of magnitude, and regardless of the magma type. We assume that the thermal conductivity λ equals 0.9 W m$^{-1}$ K$^{-1}$ (de Kleer and de Pater, 2017). By using Fourier's equation, the local heat flux density, Q, is:

$$Q = -\lambda \, \Delta T / \Delta x$$

$\Delta T$ is the temperature difference and $\Delta x$ is the thickness. We get that:

$$\Delta x = -\lambda \, \Delta T / (Q/S) = -\lambda \; \Delta T / M_e$$

where $M_e$ is the radiant emittance. From which it follows that in all three cases, the thickness is between 5 and 10 meters. In the fourth case, since virtually the emittance has no lower limit (it is not measurable), the crust could be much thicker. This estimation of thickness can be useful to estimate the average crustal overturning rate, that is, how long the present crust has existed without being overtopped by new lava or sinking: according to de Kleer and de Pater (2017), 5 meters is achieved after one year, and 10 meters after a few years. A thick crust would also help explain why there is no radiance caused by cracks or fountains observed in the center of the lava lake, as we see in the thinner-crust lakes at Puʻu ʻOʻo in Hawaiʻi (Figure S1 in supplementary material) and Erta Ale.

## The thermal infrared emission from lava lakes

As anticipated, most of the observed lava lakes have a ring of brighter emission close to the containing patera edge. This was interpreted in a previous work (Mura et al., 2024b) as the evidence for up-and-down movement of the whole lake surface. The alternative explanation given in Mura et al. (2024b) is that the pattern is due to convection in the magma, but this would imply that the crust should show some variability in the radiance (due to variability in the thickness) which we do not see (instead, we sometimes see abrupt changes in a very few cases, like Pfu1063). At least for the four cases discussed in the previous section, the thickness of the crust is estimated to be at least one year old, with no apparent variation in crustal thickness from the center to the perimeter. Even if the lava lake was convecting, since those lakes have a radius of 20 or 30 km, the upper limit for the horizontal velocity in a convection cell would be ~5 cm/day. This is a very low value that would make the convection hypothesis implausible (see Mura et al., 2024b, Figure 6, panels A and B).

Mura et al. (2024b, figure 4) showed the radiance variation at different observation points for four Chors Patera observations. The lava ring's structure remains unchanged, but the peak brightness shifts based on the observer's position. This peak always appears opposite to the observer, likely because the patera floor is significantly lower than the surrounding area, with the lava ring near the edges and partly obscured by steep, possibly vertical walls. Shadow calculations made by Radebaugh et al. (2001) suggest these walls can be up to 900 meters high. This is consistent with the idea that a piston up-and-down movement of the lake surface would not spill lava outside the patera very often (indeed, we do see few flows outside of paterae, but in most cases we do not see any). This is also consistent with the observation (Mura et al., 2024a) that the radiance from hotspots goes down with the emission angle ($e$) faster than the usual cos($e$), just as would be expected if the hotspot is well below the rim of the patera.

As previously reported, the apparent annular width (Mura et al., 2024b) is larger than the actual annular width, because these features are sub-pixel. To estimate the actual annular width, we repeat the original calculation with the larger database we now have. The emissivity of lava increases as temperature decreases, ranging from 0.6 at 900 K to 0.95 at 500 K (Ramsey et al., 2019); we consider two possible cases here. Case *A* is a black body with a temperature of ~900 K (emissivity 0.6), which emits a radiance of ~500 W sr$^{-1}$ m$^{-2}$ in the 4.5-5.0 μm range; case *B* is a black body with a temperature of ~500 K and emissivity of 0.95, which emits a radiance of ~50 W sr$^{-1}$ m$^{-2}$ in the same range. For the L band (3.3 to 3.6 μm), the uncertainty is larger, because the two scenarios lead to 400 and 20 W sr$^{-1}$ m$^{-2}$ in cases A and B (they differ by a factor 20). Also, typically the signal is lower in the L band, so that the signal to noise ratio is worse. However, since we have many cases, we can attempt a systematic retrieval of the actual annular width.

The result is in Figure 4; the distributions of apparent annular width have very long tails, so we prefer to indicate the median value (3 m or 40 m depending on the model) and not the average. For the same reason, the uncertainty is not calculated as the standard deviation, but taking the FWHM of the two distributions; hence, the results for the two model can be expressed as $3^{+5}_{-2}$ and $40^{+60}_{-30}$ meters.

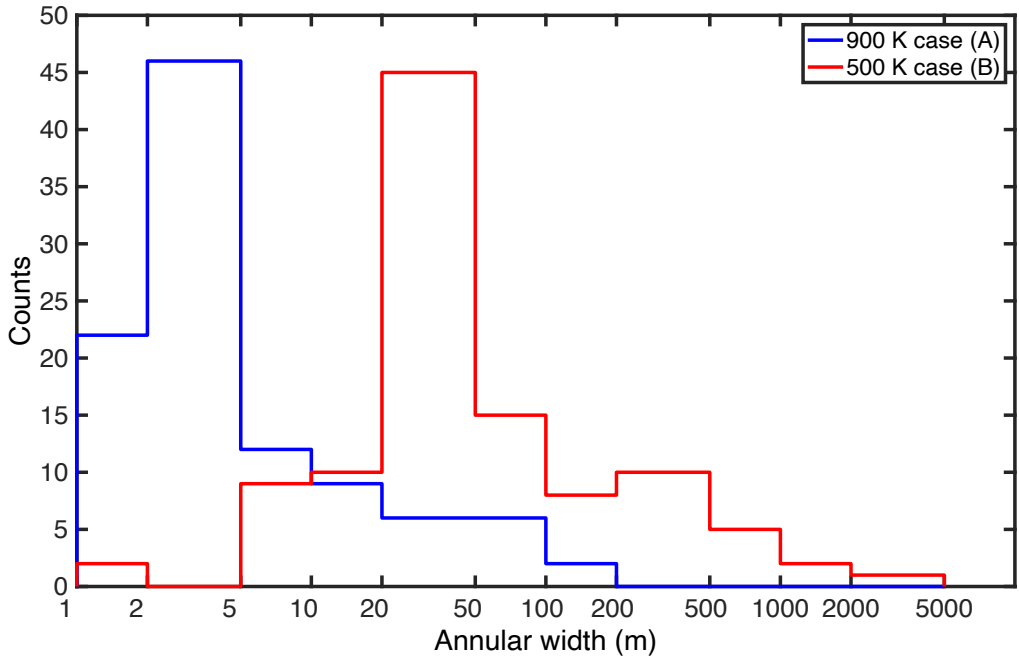

**Figure 4.** Distribution of estimated annular widths (for all of the observations, including multiple observations of the same patera at different times) in the two scenarios: 900 K (A) and 500 K (B). The median values are respectively 3 and 40 meters; with uncertainties, the values, in meters, are $3^{+5}_{-2}$ and $40^{+60}_{-30}$. Counts are not normalized by the bin size; bins edges are indicated as the x axis labels.

In this very simplified model for a lava lake, the M-band power output is easy to calculate. If we schematize it as an annulus of perimeter $p$ and width $d$, the M-band power output $P$ is just $p\,d\,R$, where $R$ is the radiance. In the two cases $A$ and $B$ the product of $d$ and $R$ is similar (this is not by chance, but a result of the way $d$ is calculated from $R$) and has the value of $20^{+30}_{-15}\,10^2$ W sr$^{-1}$ m$^{-1}$. This is of course a very rough estimation, and one could expect large deviations from this relationship, because in many cases the emission is not coming from a regular, complete ring, nor is it coming from features close to a ring, or the lava lake is a boiling or Chaotic type. However, we note that the data falls quite well within the model limits (Figure 5). Even if the data is quite scattered, the correlation coefficient between $P$ and $p$, (0.40) is well above the Student-test correlation (i.e. the critical value for Pearson's correlation coefficient) for a sample of ~50 points and a threshold of 5% (0.27).

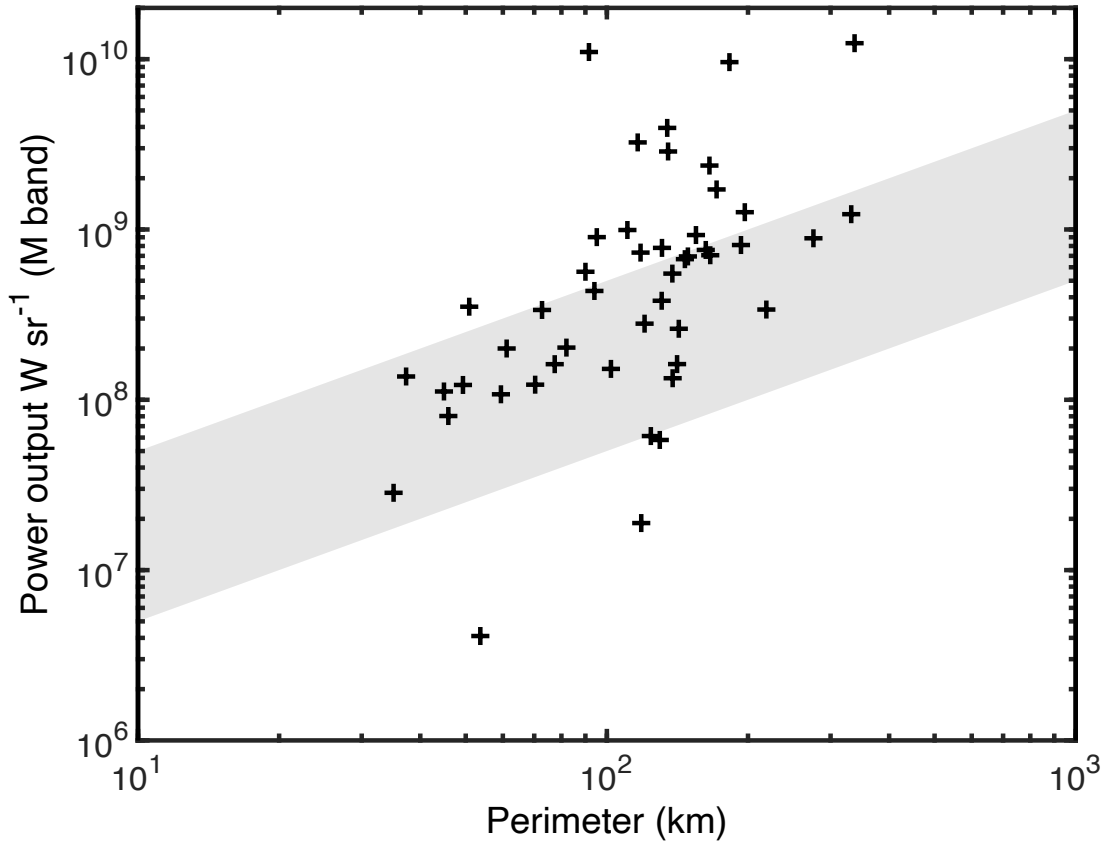

**Figure 5.** Scatterplot of the M-band power output of lava lakes as a function of their perimeter. The boundaries of the model ($P= p\,d\,R$, assuming has the value of $20^{+30}_{-15}$) are indicated by the gray area (that is, the top margin of the gray area is when one assumes that $d{\cdot}R$=5 $10^2$ W sr$^{-1}$ m$^{-1}$, the bottom one is for $d{\cdot}R$=50 $10^2$ W sr$^{-1}$ m$^{-1}$.

We again note that such model is a simplification, and it does not mean that the thermal emission is restricted to annuli with a diameter of several km, and with an annular width of a few meters, where the lava is uniformly exposed, close to an area where the crust uniformly covers the magma. It is plausible that there is a mixed zone. Therefore, the thickness we give should be interpreted as equivalent to that of a ring of completely exposed lava. In any case, even if we relax -because of the previous consideration - this few-meter value by a factor 10 or even 100, the vast majority (almost all) of the lake is occupied by crust. We can therefore calculate the implication of this in terms of thermal radiation; in particular, we want to evaluate how much of the heat is radiating from the lava and how much from the crust.

Chors Patera is a case that has a good resolution and that clearly shows a uniform crust occupying most of the patera. Below are some numbers from the best observations, that is the one from orbit 55. The total lake surface is ~2200 km$^2$, the perimeter is ~170 km. The total M-band power output, as observed by JIRAM, is 1 GW. The average M-band radiance from the crust is ~40 mW sr$^{-1}$ m$^{-2}$, which, integrated over the lake surface (that's a fair assumption since the crust occupies most of it) gives a value of 0.25 GW. In summary, in the M band, only 25% of the output comes from the crust.

Now, let us assume that Chors is modelled as a hotspot where most of output comes from a body at either 500 or 1000 K. At 500 K the radiance of a black body, integrated in the M band, is 1/20 of the total; at 1000 K is 1/15 of the total. One would then extrapolate that the total power output (not completely accessible from near-IR observations) of Chors is 15 or 20 GW. Note that a similar value could be obtained if we take the area of the ring (actual annular width of the ring by the perimeter) and multiply it by the emittance of a black body at 500 K or 1000 K.

However, the crust radiance, 40 mW sr$^{-1}$ m$^{-2}$, is indicative of a temperature between 225 and 230 K. Integrating the corresponding black-body emittance over the lake's surface, we obtain a value between 250 and 300 GW. In practice, using the M band and assuming it as a proxy for the total power output may lead to an underestimation of the thermal output up to a factor of 10 to 20 for those hot spots where the spatial resolution is not enough to give the detailed temperature map as in this case.

While the actual numbers of this underestimation may vary for other lava lakes (for example: Catha looks similar to Chors; Pfu1063 has a uniform crust radiance of about 30 mW sr$^{-1}$ m$^{-2}$ in a large part of patera, but there is also a dark part; U07 shows a dark and plausibly low-emitting crust), we have evidence, from the distribution of the annular width in Figure 3, that many lava lakes have similar annular width and hence most of the lake's surface is covered by crust.

This analysis does not consider the potential presence of multiple tiny fractures within the crust that are too small to detect. Such hot cracks are common features on active terrestrial lava lakes (Flynn et al., 1993), but in such instances it is the high overturn rate that keeps the crust sufficiently thin to allow the crust to break more easily. If such fractures are present and evenly distributed on Io, the crust could be thicker than our calculations indicate, so that the cracks contribute more to the thermal output than radiance through the crust. While there is some evidence suggesting that the crust may have fractures, (see for example Pfu1063 in Figure 1) these are typically large and clearly visible. In fact, Matson et al. (2006) excluded these cracks from their thermal model, because they suggest that their overall impact was too minor to be of significance. Also, the crustal temperatures we estimate are in good agreement with those from de Kleer and de Pater (2017), which would not be the case if substantial contribution of the radiance comes from very hot cracks.

Hence, the bulk of the power output from a lava lake, at least in many cases, comes from the crust and not from the exposed lava at the edges; the total output inferred from the near infrared spectrum can be misleading (unless thermal infrared data is available, covering the whole crust radiance that peaks at wavelengths larger than 10 μm).

## Summary and conclusions

While previous studies noted the presence of likely lava lakes (Radebaugh et al., 2001; Lopes et al., 2004), the spatial resolution precluded observation of the thermal structure of these features, except in glimpses of curvilinear edges in high-resolution Galileo images and enhanced thermal emission around the edges of a few paterae in NIMS images. The JIRAM IR observations of hot rings at the edges of Io's paterae have been a revelation about the style of volcanic eruption there. Visible images of paterae reveal that they have thin, dark, margins that are thought to result from filling and draining of the patera. Bathtub rings in terrestrial volcanic craters, such as the 1959 Kilauea Iki lava lake in Hawaii, probably form from similar processes (Stovall et al., 2008). It is notable that on Earth, long-lived, persistent lava lakes are rare, yet they appear to be a major volcanic feature on Io: it appears that if imaged at the right time and with high enough spatial resolution, we might see hot rings at most active paterae, which also mostly appear to be lava lakes, on unique Io.

Using JIRAM observations made during orbits 51 – 62, we have identified at least 40 paterae on Io with thermal characteristics believed to be associated with active lava lakes. This implies that active lava lakes are ubiquitous on Io. The annular width of the spattering region around the margins of the patera floors is of the order of a few meters to tens of meters, while the diameter of the observed lava lakes ranges from 10 to 100 km. Most interesting is the observation that the hot perimeter of each patera floor is almost always continuous. Whatever mechanism creates this hot perimeter must operate at all azimuths irrespective of the patera diameter. A central upwelling of fresh magma is unlikely to produce this uniform characteristic of the perimeter.

We have been able to estimate the thickness of the crust in the center of some lava lakes to be of the order of 5-10 m, suggesting that it is a few years old. The bulk of the thermal emission comes from the crust and not from the exposed lava, so the total power output cannot be calculated from the 5-μm radiance alone. These attributes have important implications for the magma production rate at each volcanic source, as well as the mechanism for the overturning of the lava lake. As we proposed in our earlier investigation (Mura et al., 2024b) the motion of the lake surface is most consistent with a piston-like motion of the entire floor, rather than a resurfacing from the center of the lake. The cause of this piston motion remains to be resolved, but conceptually there may be some tidal action producing vertical motions in a lake surface as Io orbits Jupiter (de Kleer et al., 2019). The fact that so many patera simultaneously illustrate the same phenomenon at widely spaced places on the moon would argue for the tidal model. The theoretical basis for this model does, however, await development. While a few patera (e.g., P62, Amaterasu, Unnamed 18, and Tol Ava patera) are bright across the entire lake, such thermally dynamic features are rare. We interpret these thermal anomalies as demonstrating that some lava lakes experience vigorous overturning. Connecting the process responsible for vigorous overturning to the more quiescent state also needs further investigation.

We note that eight previously unidentified hotspots have been found (two of them, Dingir P. and Hatchawa P., were known as paterae but not reported as active hotspots); this analysis is restricted to those that show a peculiar morphology, hence it is expected that future analysis of this dataset will reveal more unidentified hotspots. This raises the question of how many additional patera may be active at one time, which is important for the estimation of the global heat flux (Hamilton et al., 2013); this might be resolution-dependent too. Lakes with completely dark crust, a characteristic of very low temperatures and very thick and old lavas have also been observed. Such features may be lakes that have a highly variable periodicity.

It is also worth considering what observations, to be made from some future space mission yet to be approved, might help positively identify the nature of the mechanism that produces the bright rings. One such measurement would be a higher spatial resolution view of the thickness variations across each of the patera. If there were sufficient spatial resolution to resolve differences in crustal thickness within a lava lake, then the resurfacing mechanism would be easier to understand. If the lake had thinner crust near the center than at the margins, then this would favor a resurfacing mechanism operating from central upwelling. But if the crust has uniformly the same thickness, then a piston-like mechanism seems more reasonable. Finally, it would also be instructive to compare crust thickness at different patera. It seems reasonable to expect paterae with the highest heat flow would have the thinnest crust, so that a global survey of crustal thicknesses at paterae might provide new insights into the global distribution of heat flow.

## Acknowledgments

We thank Agenzia Spaziale Italiana (ASI) for the support of the JIRAM contribution to the
Juno mission. This work is funded by the ASI–INAF Addendum n. 2016-23-H.3-2023 to grant
2016-23-H.0. Part of this work was performed at the Jet Propulsion Laboratory, California
Institute of Technology, under contract with NASA.

## Open Research

The JIRAM dataset used for our analysis is publicly available at the Juno Archive at the
Planetary Atmospheres Node https://pds-
atmospheres.nmsu.edu/PDS/data/PDS4/juno_jiram_bundle/data_calibrated/

The Voyager-Gaileo data used for our analysis is available through Williams et al., 2011b

## Corresponding author

Correspondence to Alessandro Mura (alessandro.mura@inaf.it)

## Competing interests

The authors declare no competing interests.

## Methods

**Geometry reconstruction.** We utilize the NAIF-SPICE software (Acton, 1996) to project images onto Io's surface. For each orbit, we select the highest quality and spatial resolution images from the dataset, resulting in the use of fewer than the original 2000 images. To enhance the accuracy of the geometric reconstruction, we verify and adjust the coalignment of these images as needed. To minimize noise, we stack the coaligned images to create a 3D data cube, then compute the median value along the third dimension to produce a single 2D super-image per orbit and band (following the method of Mura et al., 2020, 2024b). After this process, we estimate the uncertainty in the geometric reconstruction of JIRAM images to be no more than one pixel. Since the de-spinning mirror of JIRAM was inoperative during the observations, we employed very short integration times (2 ms on average) to minimize smearing (0.8 pixels/ms). Residual smearing was corrected using a Lucy-Richardson algorithm (Lucy, 1974; implemented in MATLAB).

**Estimation of size and circularity.** To calculate the area of the hotspots that show a ring morphology, we attempted the reconstruction of the shape of the ring by means of a simple automatic pattern recognition algorithm. The purpose of this calculation is also to quantify how much it deviates from a circular shape. Once an approximate center point of the feature is located, the algorithm finds the "edges" of the ring by calculating the maximum along a radius from the center outward. The location of these maxima $r_i$ (one for each direction of the radius) are then fitted with a function that gives $r$ as a function of the angle $\theta$ and has this equation:

$$r = r_0 + \sum_{j=1}^{n} \left( c_j \sin(\theta) + d_j \cos(\theta) \right)$$

where $r_0$, $c_j$ and $d_j$ are free parameters, and n is chosen between 1 and 3 depending on the complexity of the feature. Once the contour of the "ring" is found, the exact center can be calculated. Finally, a circularity factor $f$ is calculated as:

$$f = \sqrt{1 - \left( \frac{min(r)}{max(r)} \right)^2}$$

$f$ is 0 for a perfect circle (it's the analogous of the eccentricity for an ellipse).